\documentclass[aps,prl,reprint]{revtex4-1}

\usepackage{amsmath,amssymb,graphicx,color,verbatim,soul}
\usepackage{upgreek}
\usepackage{graphicx}
\usepackage{float}
\usepackage{hyperref}
\hypersetup{colorlinks=false}

\begin{document}

\title{Simultaneous Readout of Noncommuting Collective Spin Observables beyond the Standard Quantum Limit}

\author{Philipp Kunkel}
\email{fullspinreadout@matterwave.de}
\author{Maximilian Pr\"ufer}
\author{Stefan Lannig}
\author{Rodrigo Rosa-Medina}
\author{Alexis Bonnin}
\author{Martin G\"arttner}
\author{Helmut Strobel}
\author{Markus K.\ Oberthaler}
\affiliation{Kirchhoff-Institut f\"ur Physik, Universit\"at Heidelberg, Im Neuenheimer Feld 227, 69120 Heidelberg, Germany}

\date{\today}

\begin{abstract}
	We augment the information extractable from a single absorption image of a spinor Bose-Einstein condensate by coupling to initially empty auxiliary hyperfine states.
	Performing unitary transformations in both, the original and auxiliary hyperfine manifold, enables the simultaneous measurement of multiple spin-1 observables.
	We apply this scheme to an elongated atomic cloud of $ ^{87} $Rb to simultaneously read out three orthogonal spin directions and with that directly access the spatial spin structure.
	The readout even allows the extraction of quantum correlations which we demonstrate by detecting spin nematic squeezing without state tomography.
\end{abstract}

\maketitle

Ultracold atomic systems have proven to be a powerful platform for implementing quantum technologies such as quantum simulation~\cite{Bloch2012} and quantum enhanced sensing~\cite{Pezze2018}.
For all experimental implementations efficient readout is essential to extract the properties of interest.
In fact, advances in readout techniques have paved the way to new discoveries. 
This includes absorption imaging to observe Bose-Einstein condensation~\cite{Anderson1995,Davis1995},
the quantum gas microscope uncovering spatial correlations in Hubbard models~\cite{Gross2017}
and dispersive methods to observe spin textures in spinor BECs~\cite{Sadler2006}.

\begin{figure}[H]
	\centering
	\includegraphics[width=\columnwidth]{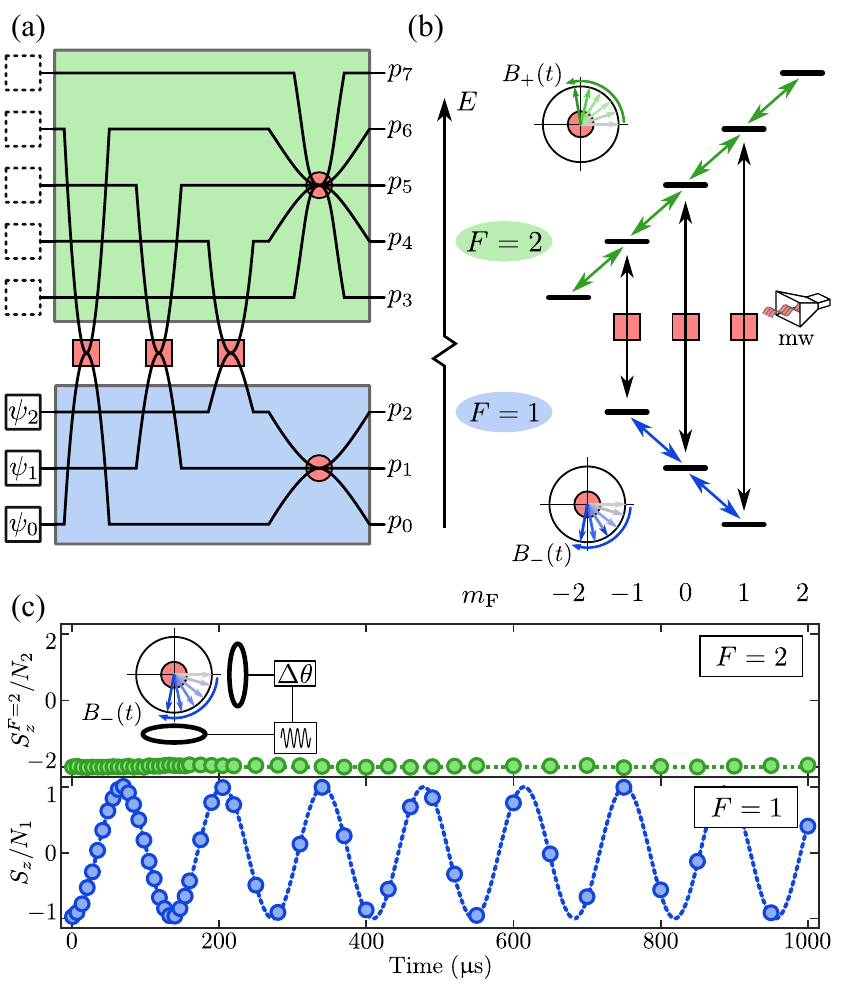}
	\caption{Schematics of the readout technique.
		(a) Coupling (red squares) the system (blue) to auxiliary states (green) enlarges the Hilbert space.
		Applying independent unitaries in both subspaces (red circles) allows choosing the measurement bases individually.
		From the measured populations ($ p_0 $,...,$ p_7 $) one infers the corresponding observables of the original system.
		(b) Level scheme of the electronic groundstate of $ ^{87} $Rb in a magnetic field. Coupling the two manifolds with mw pulses we experimentally realize the extension of the Hilbert space.
		Rotating magnetic fields selectively couple the magnetic substates in each manifold. 
		(c) Using two rf coils we generate a rotating magnetic field.
		For a relative phase of $ \Delta\theta = -0.7\,\pi$ we induce spin rotations only in $ F=1 $~\cite{SupplementaryMaterial}. For this we measure the $ z $-projections $ S_z $ and $ S_z^{F=2} $ of the spin normalized to the atom numbers $ N_1 $ and $ N_2 $ detected in the $ F=1 $ and $ F=2 $ manifold, respectively.
	}
	\label{Figure1}
\end{figure}

Here, we show a readout technique to simultaneously access noncommuting spin-1 observables and detect quantum correlations such as coherent spin squeezing.
For this we couple the original system to a set of auxiliary states which, combined with unitary transformations, enables the simultaneous readout by projective measurements of all populations in the enlarged Hilbert space~\cite{Sosa2017} (see Fig.~\ref{Figure1}(a)).
Our readout is especially advantageous in systems with additional spatial degrees of freedom.
There, a measurement in a single global basis setting for each experimental realization may not be sufficient to capture all relevant aspects of the quantum state.
A prime example is the cluster state, a valuable resource for measurement based quantum computing~\cite{Raussendorf2003}, which features spatial correlations between noncommuting observables~\cite{Loock2007}.

For demonstration, we realize our technique in a spinor Bose-Einstein condensate (BEC) of $ ^{87} $Rb in the $ F=1 $ hyperfine manifold.
The initially unoccupied $ F=2 $ hyperfine states serve as the auxiliary states to which we couple via microwave (mw) pulses (see Fig.~\ref{Figure1}(b)).
In order to selectively couple the magnetic sublevels in the two manifolds we use two orthogonal radiofrequency (rf) coils which generate a rotating magnetic field~\cite{Smith2013,Bharath2018}.
This exploits the different signs of the corresponding magnetic moments to independently induce spin rotations (see Fig.~\ref{Figure1}(c)).
Together with mw coupling between the manifolds this gives full control over the measurement basis~\cite{Smith2013} and in principle allows the simultaneous measurement of 7 spin-1 observables out of the 8 needed to completely describe a single particle state~\cite{Flammia2005}.
Such a readout scheme constitutes a generalized measurement where the formalism of positive operator valued measures (POVMs)~\cite{Peres2006} allows relating the measured populations to the expectation value of spin operators acting on the original system.

To demonstrate the possibility to spatially resolve a complex spin structure in a single realization with this readout, we prepare an elongated BEC of $ \approx 40,000 $ atoms in a dipole trap  with trapping frequencies $ (\omega_y,\omega_\perp) = 2\pi\cdot(2.3,170)\, $Hz.
All atoms are initialized in the state $ (F,m_\text{F}) = (1,-1) $ in a magnetic field of $ B=0.884\, $G along the $ z $-direction.
Using spin-rotations induced by the rf coils and a magnetic field gradient along the longitudinal direction of the BEC we generate a spin wave involving the three spin directions $ \hat S_x $, $ \hat S_y $ and $ \hat S_z $ (see supplemental material (SM)~\cite{SupplementaryMaterial} for details).
To read out all three spin directions in a single experimental realization we use the following scheme.
We first apply three mw pulses coupling $ (1,j) \leftrightarrow (2,j) $ ($ j=0,\pm1 $) to split the state between the $ F=1 $ and $ F=2 $ manifold.
Selective $ \pi/2 $ spin-rotation in $ F=1 $ around the $ y $-axis maps the spin observable $ \hat{S}_x $ onto the populations ($ p_0 $, $ p_1 $, $ p_2 $).
In order to extract $ \hat{S}_y $ as well as $ \hat{S}_z $  we apply a $ \pi/4 $ spin-rotation around the $ x $-axis in the $ F=2  $ manifold.
We ensure the phase coherence of all these pulses by active magnetic field stabilization and GPS locking of the rf and mw sources.

\begin{figure}
	\centering
	\includegraphics[width=\columnwidth]{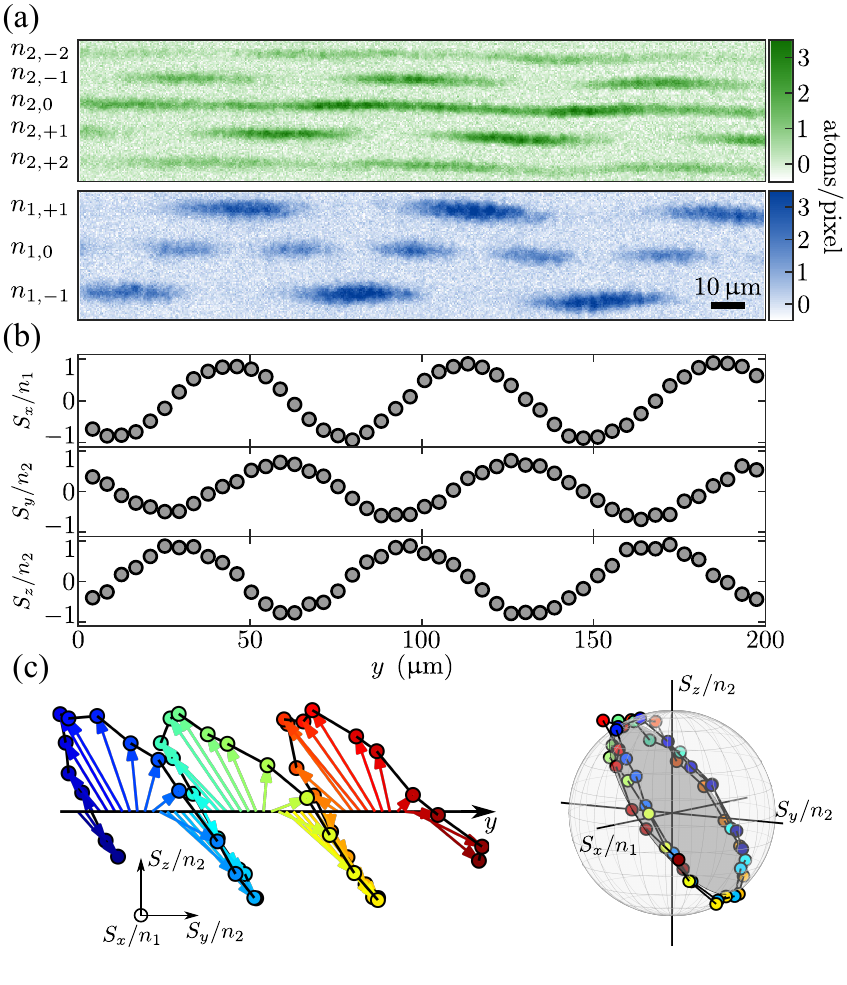}
	\caption{Spatially resolved readout of three spin components.
		Using magnetic field gradients and spin rotations we generate a spin wave along the longitudinal direction of an elongated BEC.
		(a) Populations of the magnetic substates measured via absorption imaging after Stern-Gerlach separation.
		(b)	All three spin directions inferred from the measured populations at each position. The spin observables are normalized to the local atom number $ n_F(y) $ in the corresponding hyperfine manifold.
		(c) Reconstructed spin vector in space and its distribution on a spin sphere.}
	\label{Figure2}
\end{figure}

With a Stern-Gerlach pulse we spatially separate the different $ m_\text{F} $ states and use hyperfine selective absorption imaging to measure the population in all magnetic substates with a spatial resolution of $ \approx 1.2\,\upmu $m as shown in Fig.~\ref{Figure2}(a).
After this sequence the three spin directions are extracted from the measured atom numbers as follows
\begin{equation}
\begin{alignedat}{1}
S_x(y) = & \langle\hat{S}_x(y)\rangle_{\delta y} =n_{1,+1}(y) - n_{1,-1}(y) \\
S_y(y) = &\frac{4}{\sqrt{6}}\,(n_{2,+1}(y)-n_{2,-1}(y))\\
S_z(y) = &\sqrt{2}\,[2n_{2,+2}(y)-n_{2,+1}(y) \\
&+ n_{2,-1}(y)-2n_{2,-2}(y)],
\end{alignedat}
\end{equation}
where $ n_{F,m}(y) $ is the local atom number in the evaluation interval of $\delta y \approx 5\,\upmu $m in the state $ (F,m) $.
Here, $ \langle \cdot\rangle_{\delta y} $ denotes the local mean corresponding to an average over $ \approx 700 $ particles.
This measurement yields at every position the three components of the collective spin-vector from which we reconstruct the spin wave as shown in Fig.~\ref{Figure2}(b) and (c).
\begin{figure*}
	\centering
	\includegraphics[width = \textwidth]{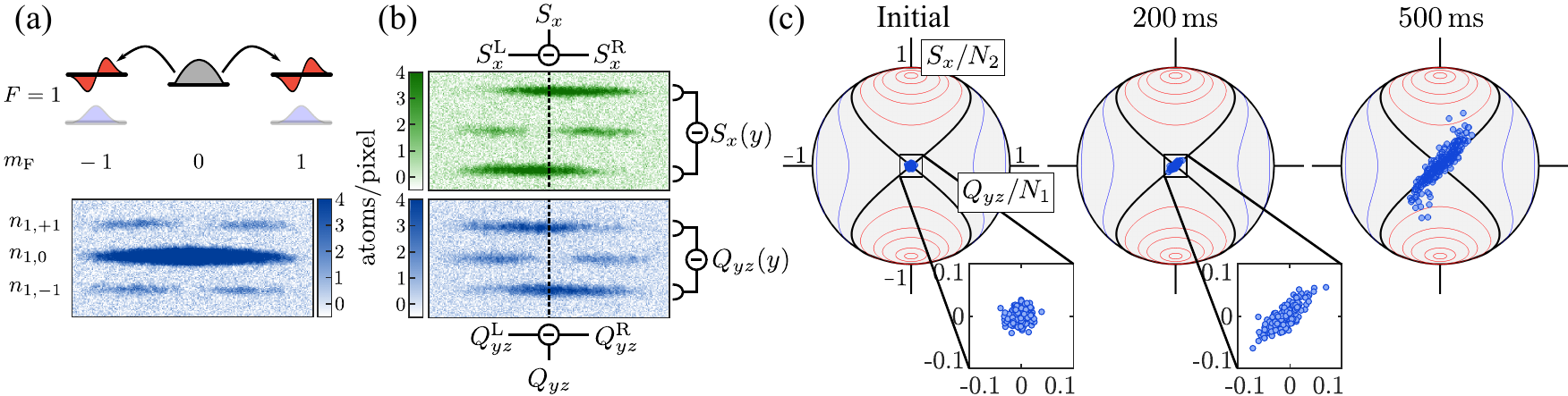}
	\caption{Efficient detection of spin mixing dynamics.
		(a) Using off-resonant mw dressing we tune spin mixing into resonance between the ground mode of $ (1,0) $ (gray) and the first excited spatial mode of $ (1,\pm1) $ (red).
		The lower panel shows an absorption image of the density distributions after 2\,s of spin mixing evolution time where the populations in $ (1,\pm1) $ feature the characteristic double peak structure.
		(b)	After the readout sequence the observables  $ S_x $ and $ Q_{yz} $ are extracted from the population differences of the states $ (2,\pm2) $ and $ (1,\pm1) $, respectively.
		Here the absorption images have been taken after 800\,ms of evolution time for better visibility of the mode structure.
		(c)	By plotting the two values for each experimental realization we directly visualize the spin-mixing dynamics in the spin-nematic phase space. 
		The lines correspond to the mean field energy contours of the phase space.
		For an evolution time of $ 500 \, $ms a non-Gaussian shape has emerged.
	}
	\label{Figure3}
\end{figure*}

In order to benchmark the capabilities of our readout scheme to extract quantum correlations we prepare an entangled state in our spin-1 system using spin mixing.
The resulting spin-nematic squeezed state features correlated fluctuations in two noncommuting observables $ \hat S_x $ and $ \hat Q_{yz} $~\cite{Hamley2012}.
Here, $ \hat{Q}_{yz} $ is a so-called quadrupole operator which captures an additional degree of freedom inherent to a spin-1 system~\cite{SupplementaryMaterial}. 

In order to constrain the dynamics to the spin degree of freedom we change the trap geometry for this experiment to $ (\omega_y,\omega_\perp) = 2\pi\cdot(40,170)\, $Hz by confining the atomic cloud with an additional crossed dipole beam.
We prepare $ \approx 20.000 $ atoms in the state $ (1,0) $ in the spatially symmetric ground state mode.
Spin mixing leads to pairwise creation of particles in the states $(1,\pm1) $ and the energy of $(1,0) $ is tuned such that this process is in resonance  with the first excited spatial mode of the effective external potential for $ (1,\pm1) $~\cite{Scherer2010}.
This mode is spatially antisymmetric which leads to the characteristic double-peak structure of the density in the states $ (1,\pm1) $ (see Fig.~\ref{Figure3}(a)).
This feature combined with our spatially resolved readout allows the implementation of common mode technical noise rejection as detailed below.
To facilitate the absorption imaging we switch off the crossed dipole beam and let the atomic cloud expand in the remaining $ \omega_y = 2\,\pi\cdot 2.3\, $Hz trapping potential for 10\,ms.

For a simultaneous readout of both observables, $ \hat S_x $ and $ \hat Q_{yz} $, we implement the following scheme. With an rf $ \pi/2 $ spin-rotation around the $ y $-direction we map the observable $ \hat{S}_x $ on the population difference of the states $ (1,\pm1) $.
We then use three mw $ \pi/2 $-pulses coupling the states $ (1,0/\pm1) $ with $ (2,0/\pm2) $ to transfer half of the population to the $ F=2 $ manifold.
In order to extract $ \hat{Q}_{yz} $ we first rotate the state back using an additional rf $ \pi/2 $ spin-rotation around the $ y $-axis in ${ F=1}$.
At this stage a spin echo sequence is used to cancel the effect of fluctuations in the magnetic field.
We then imprint a phase of $ \pi/2 $ on the state $ (1,0) $ by applying two resonant mw $ \pi $-pulses coupling the states $ (1,0) \leftrightarrow (2,0) $ with a relative phase of $ \pi/2 $.
An additional rf $ \pi/2 $-rotation then maps the observable $ \hat{Q}_{yz} $ onto the population difference of $ (1,\pm1) $ (see~\cite{SupplementaryMaterial} for a graphical illustration of this scheme).

Since the structure of the first excited spatial mode features an opposite sign between left (L) and right (R) half of the atomic cloud (see Fig.~\ref{Figure3}(b)) we evaluate
\begin{align}
\begin{split}
S_x &= S^\text{L}_x - S^\text{R}_x\\
Q_{yz} &= Q^\text{L}_{yz} - Q^\text{R}_{yz},
\end{split}
\end{align}
with
\begin{align}
\begin{split}
S^\text{L/R}_x &= \left(n_{2,+2}^\text{L/R} - n_{2,-2}^\text{L/R}\right)\\
Q^\text{L/R}_{yz} &=\left(n_{1,+1}^\text{L/R} - n_{1,-1}^\text{L/R}\right).
\end{split}
\end{align}
This analysis has the additional benefit that it mitigates fluctuations which are homogeneous over the atomic cloud such as technical noise induced by the mw and rf pulses.

For each experimental realization we obtain a point with coordinates $ S_x $ and $ Q_{yz} $ in the spin-nematic phase space and thus efficiently get an insight into the spin mixing dynamics.
In Fig.~{\ref{Figure3}}(c) we show the result for an initial state (1,0), corresponding to the preparation at the unstable fixed point of this phase space, after different evolution times.
The state expands along one axis of the separatrix (black line).
For longer evolution times $ \gtrsim 500\, $ms the state clearly becomes non-Gaussian which is directly captured with our readout without state reconstruction.
Here, we use only $ \sim $300 experimental realizations to reveal this feature.

\begin{figure}
	\centering
	\includegraphics[width = \columnwidth]{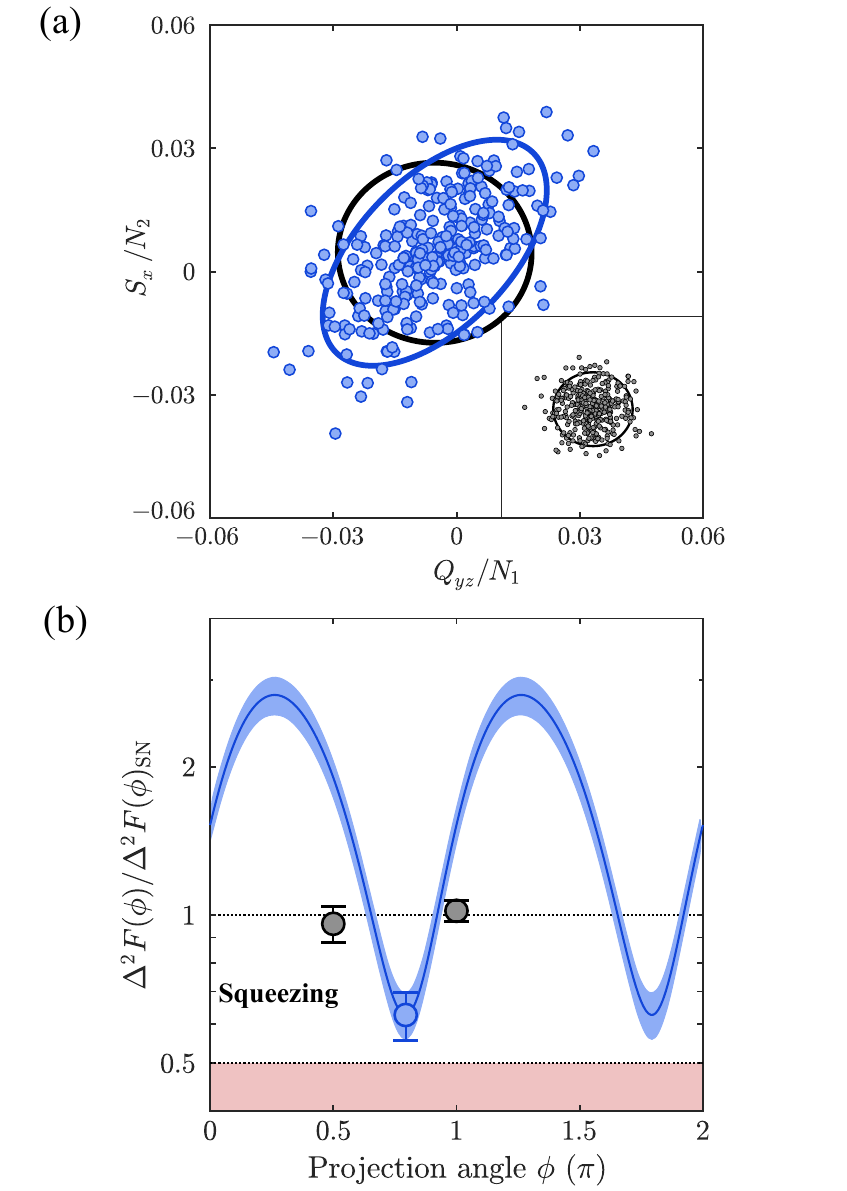}
	\caption{Detection of spin-nematic squeezing.
		(a) In each experimental run we obtain a pair of values $ S_x/N_2 $ and $ Q_{yz}/N_1 $.
		After 100\,ms of evolution time the resulting scatter plot clearly indicates correlated fluctuations between the two observables compared to the initial state (inset).
		The blue and black line depict the $ 2 $ standard deviation (s.d.) interval.
		(b) For each projection axis parameterized by the angle $ \phi $ we compute the variance. We infer $ \Delta^2 F(\phi) $ by subtracting imaging noise and normalize to the expected coherent state fluctuations $ \Delta^2 F(\phi)_\text{SN} $ resulting in the blue line and 1\,s.d. error band.
		The prepared state shows reduced fluctuations of $ 0.62\pm0.07 $ (blue point) along the maximally squeezed direction. The red shading indicates the fundamental limit of 0.5 of the readout scheme.
		The two gray points correspond to the imaging calibration of the $ F=1 $ and $ F=2 $ manifolds using a coherent spin state.
		The error bars correspond to the $ 1$\,s.d. interval.}
	\label{Figure4}
\end{figure}

For the short time dynamics one expects to find spin-nematic squeezing below the initial coherent state fluctuations indicating the creation of an entangled many-body state~\cite{Hamley2012}.
In Fig.~\ref{Figure4}(a) we plot the values of $ S_x/{N_2} $ vs. $ Q_{yz}/{N_1} $ normalized by the total atom numbers $ N_F $ measured in the hyperfine manifold $ F=1,2 $ after an evolution time of $ 100\, $ms (blue points).
The squeezing, i.e. the reduction of fluctuations along one direction at the cost of enhanced fluctuations along the orthogonal direction, is apparent.
For a quantitative ana-lysis, we compute the variance $ \Delta^2 F(\phi) $ with $ F(\phi) = \cos(\phi)Q_{yz}+\sin(\phi)S_x $. Calculating the corresponding atomic shot noise from a multinomial distribution yields $ \Delta^2 F(\phi)_\text{SN} = \langle \cos^2(\phi) N_1+ \sin^2(\phi) N_2 \rangle $ with which we normalize the variance (see Fig.~\ref{Figure4}(b)).
Note that for perfect mw $ \pi/2 $-pulses this term becomes independent of the phase $ \phi $, while in our experimental realization we observe a small imbalance corresponding to $ 0.53\, \pi$-pulses.
We infer minimal fluctuations of $ 0.62 \pm 0.07 $ clearly below the standard quantum limit where independently characterized imaging noise contributions have been subtracted. Without subtraction we find a value of $ 0.81 \pm 0.07 $.
By measuring the fluctuations of a coherent spin state, we independently calibrated our imaging for $ F=2 $ and $ F=1 $ corresponding to $ \phi= 0.5 \,\pi  $ and $ \phi = \pi $, respectively~\cite{SupplementaryMaterial} (gray points in Fig.~\ref{Figure4}(b)).
After Stern-Gerlach splitting all relevant densities for extracting $ S_x $ and $Q_{yz} $ are spatially non-overlapping since the magnetic moments of $ (2,\pm2) $ are twice as large as the ones of $ (1,\pm1) $.
Thus, we extract all populations from a single exposure without the need for hyperfine selective absorption imaging which has the additional benefit of reduced imaging noise.

The noise suppression by nearly a factor of 2 (3\,dB) is close to the fundamental limit of our readout method. 
This limit results from the mw couplings to empty auxiliary states which individually act as beam splitters and thus each introduces additional binomial atom number fluctuations between its output ports. 
In the case of $ 50/50 $ beam splitters, the fluctuations that are extracted from measuring the signal in one port of each beam splitter lead to the estimated variance $ \Delta^2 F(\phi) $ which is then connected to the variance $\Delta^2 F_{\rm{in}}(\phi)$ of the input state of the beam splitters as follows:
\begin{equation}
\zeta(\phi) = \frac{\Delta^2 F(\phi)}{\Delta^2 F(\phi)_\text{SN}} = \frac{1}{2}\frac{\Delta^2 F_{\rm{in}}(\phi)}{N_{\rm{tot}}} +\frac{1}{2}
\label{EqSqueezLim}
\end{equation}
with $N_{\rm{tot}}=N_1+N_2$, see SM~\cite{SupplementaryMaterial} for details. 
Therefore the squeezing measured with this readout cannot submerge the bound of $1/2$ even for vanishing variance $ { \Delta^2 F_\text{in}(\phi) = 0}$.
From the measurement we infer minimal and maximal fluctuations of $ \zeta(\phi_\text{min}) = 0.62\pm0.07 $ and $\zeta(\phi_\text{max}) = 2.80\pm0.25  $. 
Using Eq.~\eqref{EqSqueezLim} we compute $  \Delta^2F_\text{in}(\phi_\text{min})\Delta^2F_\text{in}(\phi_\text{max}) = (1.1 \pm 0.6)N_\text{tot}^2 $ consistent with a minimal uncertainty state expected from the dynamics.

In summary, we demonstrate a new technique for the simultaneous readout of multiple spin components of a trapped atomic spinor gas. 
In situations where a complex valued order parameter arises whose spatial correlations are of interest the simultaneous determination of orthogonal spin-components can access these correlations.
For example the easy-plane ferromagnetic phase of a spinor gas is characterized by the order parameter $ S_x +iS_y $~\cite{Sadler2006,Williamson2016,Pruefer2018}.
Furthermore, our readout allows the direct extraction of phase space distributions without state tomography even below the shot noise limit revealing genuine quantum correlations. 
The ability to extract spatially resolved information about phase space distributions of a state is crucial in situations where a priori knowledge about the quantum state is missing.
Especially in situations of complex multimode dynamics our technique can assess the usefulness of the emerging states for quantum information processing applications such as one-way computation.

We thank Dan Stamper-Kurn and Daniel Linnemann for discussions and Marcus Huber for pointing out the connection to POVMs. This work was supported by the ERC Advanced Grant Horizon 2020 EntangleGen (Project-ID 694561) and the DFG Collaborative Research Center SFB1225 (ISOQUANT), by Deutsche Forschungsgemeinschaft (DFG) under Germany's Excellence Strategy EXC-2181/1 - 390900948 (the Heidelberg STRUCTURES Excellence Cluster) and the Heidelberg Center for Quantum Dynamics. P.K. acknowledges support from the Studienstiftung des deutschen Volkes.

\bibliographystyle{apsrev4-1}
\bibliography{FullSpinReadoutBib}
\clearpage
\setcounter{equation}{0}
\renewcommand\theequation{S\arabic{equation}}
\setcounter{figure}{0}
\renewcommand{\thefigure}{S\arabic{figure}}
\section{Supplemental material}
\subsection{Selective addressing of $ F=1 $ and $ F=2 $}
An important prerequisite for our readout technique is the ability to selectively induce spin rotations in both hyperfine manifolds which allows setting a different measurement basis in each of them.
In our experiment we apply a constant magnetic field of $ 0.884\, $G in $ z $-direction.
Linearly oscillating magnetic fields perpendicular to the offset field are routinely used to couple the magnetic substates, i.e. induce spin rotations, where the resonance frequency corresponds to the energy splitting of the magnetic substates with $ \Delta m_\text{F} = \pm1 $.
At our magnetic field the resonance frequency for the two hyperfine manifolds differs by $ \approx 2.5\, $kHz which is smaller than the Rabi frequencies used in our experiment.
Therefore, in terms of resonance frequencies a resonant coupling within one manifold leads to an off-resonant coupling within the other manifold.

An important difference, however, between the two manifolds is the different signs of the $ g $-factors.
Because of this the spins in the two hyperfine manifolds couple to different directions of rotation of an oscillating magnetic field.
To exploit this difference we use two rf coils at an angle of $ 90^\circ $.
Each of the two coils generates a linearly oscillating magnetic field at the position of the atoms where we matched the amplitudes by matching the respective resonant Rabi frequencies induced by each coil.

Tuning the phase $ \Delta \theta $ between the two rf fields allows the generation of a rotating magnetic field and the control its direction of rotation.
Depending on $ \Delta \theta $ the Rabi frequency in each manifold is given by
\begin{equation}
\Omega_{F} = 2\Omega_0\left|\sin\left(\frac{\Delta \theta - \theta_{0,F}}{2}\right)\right|
\label{RabiFreq}
\end{equation}
with some offset phase $ \theta_{0,F} $ for $ F=1,2 $.

\begin{figure}
	\centering
	\includegraphics[width=\columnwidth]{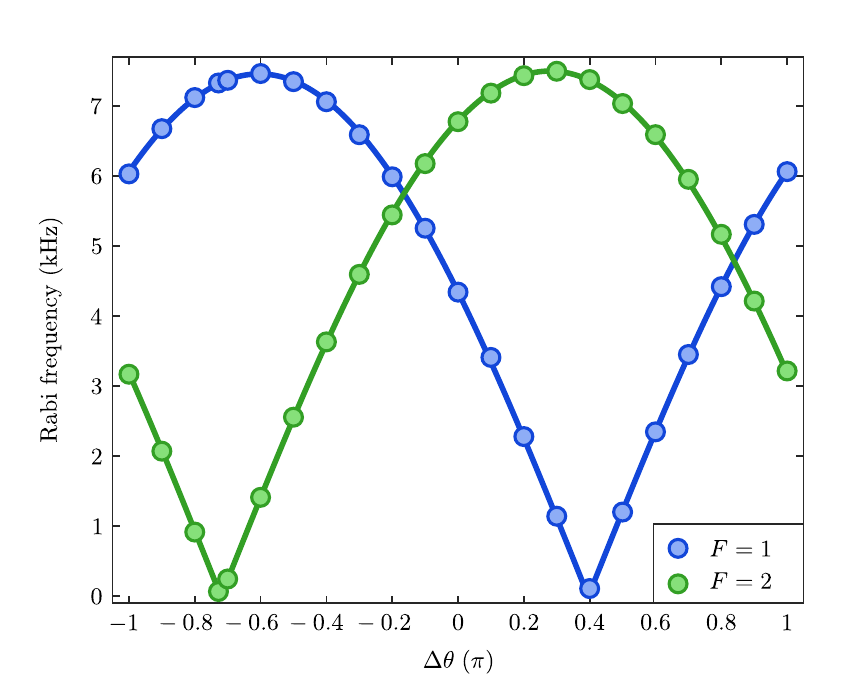}
	\caption{Measured Rabi frequency as a function of the relative phase $ \Delta \theta $ between the two rf-coils. The size of the error bars is smaller than the plot markers. We fit the function~\eqref{RabiFreq} to the experimental data and find $ \theta_{0,1}= 0.40\,\pi  $ and $ \theta_{0,2}= -0.72\,\pi  $.
		The phase mismatch between the minimal frequency in $ F=1 $ and the maximal one in $ F=2 $ (and vice versa) comes from the fact that the magnetic fields generated by the two rf coils are not perfectly orthogonal at the position of the atoms.}
	\label{SelectiveRabi}
\end{figure}

To measure these phases for both manifolds we prepare half of the atoms in $ (1,-1) $ and the other half in $ (2,-2) $ by employing a mw $ \pi/2 $-pulse coupling the two states.
We subsequently apply the rf coupling with the two rf coils where we tune the phase $ \Delta \theta $ between them.
As shown in Fig.~\ref{Figure1}(c) we record the Rabi oscillations in both manifolds and make a fit to the data to extract the resonant Rabi frequency vs $ \Delta \theta $ (see Fig.~\ref{SelectiveRabi}).
We fit the resulting curve according to Eq.~\eqref{RabiFreq}.
For orthogonal magnetic rf field vectors one can switch between a purely left- and right-rotating field by changing $ \Delta\theta $ by $\pi $.  
In our case, we find a small deviation by $\theta_{0,1} = \theta_{0,2}+1.12\,\pi $ which is consistent with the magnetic field lines being non-orthogonal but deviating by a small angle of $ 6^\circ $.
Yet, this does not obstruct our ability to fully suppress Rabi rotations in one manifold while having close to maximum coupling in the other one.

\subsection{Definition of spin-1 operators}
A general spin state in $ F=1 $ (spin-1) is {characterized} by a set of 8 generators of the su(3) Lie algebra. Here we choose the three spin operators which in second quantization are defined as
\begin{align}
\begin{split}
\hat{S}_x &=\frac{1}{\sqrt{2}}\left[\hat{a}^\dagger_{1,0}\,(\hat{a}_{1,+1}+\hat{a}_{1,-1}) \enspace + \enspace \text{h.c.}\right] \\
\hat{S}_y &=\frac{i}{\sqrt{2}}\left[\hat{a}^\dagger_{1,0}\,(\hat{a}_{1,+1}-\hat{a}_{1,-1}) \enspace + \enspace \text{h.c.}\right] \\
\hat{S}_z &=\hat{a}^\dagger_{1,+1}\hat{a}_{1,+1}-\hat{a}^\dagger_{1,-1}\hat{a}_{1,-1} \\
\end{split}
\end{align}
where $ \hat{a}^\dagger_{F,m} $ ($ \hat{a}_{F,m} $) is the creation (annihilation) operator in the state $ (F,m) $. 
Additionally, we include five quadrupole operators defined as~\cite{Carusotto2004,Hamley2012}
\begin{equation}
\begin{alignedat}{2}
&\hat{Q}_{zz} &&= \frac{2}{3}(-2\hat{a}_{1,0}^\dagger\hat{a}_{1,0}+\hat{a}_{1,+1}^\dagger\hat{a}_{1,+1}+\hat{a}_{1,-1}^\dagger\hat{a}_{1,-1})\\
&\hat{Q}_{xx} &&= \frac{1}{3}(2\hat{a}_{1,0}^\dagger\hat{a}_{1,0}-\hat{a}_{1,+1}^\dagger\hat{a}_{1,+1}-\hat{a}_{1,-1}^\dagger\hat{a}_{1,-1})\\
& && +\hat{a}_{1,+1}^\dagger\hat{a}_{1,-1}+\hat{a}_{1,-1}^\dagger\hat{a}_{1,+1}\\
&\hat{Q}_{xy} &&= i\hat{a}_{1,-1}^\dagger\hat{a}_{1,+1}+\;\text{h.c.}\\
& \hat{Q}_{xz} &&= \frac{1}{\sqrt{2}} \hat{a}_{1,0}^\dagger(\hat{a}_{1,+1} - \hat{a}_{1,-1}) + \; \text{h.c.} \\
& \hat{Q}_{yz} &&= \frac{i}{\sqrt{2}} \hat{a}_{1,0}^\dagger(\hat{a}_{1,+1} + \hat{a}_{1,-1}) + \; \text{h.c.}. 
\end{alignedat} 
\end{equation}
We use these operators as a basis set to express the result of our measurements.

\subsection{Unitary transformations}
The full readout schemes employs unitaries and projective measurements in the $ S_z $ basis in both manifolds.
Extension of the Hilbert space is implemented by mw coupling of the two hyperfine manifolds described by the operators
\begin{equation}
\begin{alignedat}{1}
\hat{C}^{ij}_{x}&= \frac{1}{\sqrt{2}}\hat{a}^\dagger_{1,i}\hat{a}_{2,j}^{\phantom{\dagger}} + \text{h.c.}\\
\hat{C}^{ij}_{y}&= \frac{i}{\sqrt{2}}\hat{a}^\dagger_{1,i}\hat{a}_{2,j}^{\phantom{\dagger}} + \text{h.c.},
\end{alignedat}
\end{equation}
which couple the states $ (1,i) \leftrightarrow (2,j) $.

Spin-rotations are implemented via rf-pulses.
In our experiment, the $ z $-direction is defined by the applied magnetic field while we define the $ y $-direction with the first rf rotation, setting the reference frame for all further manipulations.
The rotation axis for subsequent rf-pulses is defined by the relative phase with respect to the initial pulse.
The spin-rotations in $ F=2 $ are described by the two spin-2 operators
\begin{equation}
\begin{alignedat}{2}
&\hat S_x^{F=2} && = \hat{a}_{2,+1}^\dagger(\hat{a}_{2,+2}+\sqrt{\frac{3}{2}}\hat{a}_{2,0})\\
& &&+\hat{a}_{2,-1}^\dagger(\sqrt{\frac{3}{2}}\hat{a}_{2,0}+\hat{a}_{2,-2}) +\text{h.c.}\\
&\hat S_y^{F=2} && = i\hat{a}_{2,+1}^\dagger(\hat{a}_{2,+2}-\sqrt{\frac{3}{2}}\hat{a}_{2,0})\\
& &&-i\hat{a}_{2,-1}^\dagger(\sqrt{\frac{3}{2}}\hat{a}_{2,0}-\hat{a}_{2,-2}) +\text{h.c.}\,.
\end{alignedat}
\end{equation}
\subsection{Simultaneous readout of $ S_x $, $ S_y $ and $ S_z $}
\begin{figure}
	\centering
	\includegraphics[width=\columnwidth]{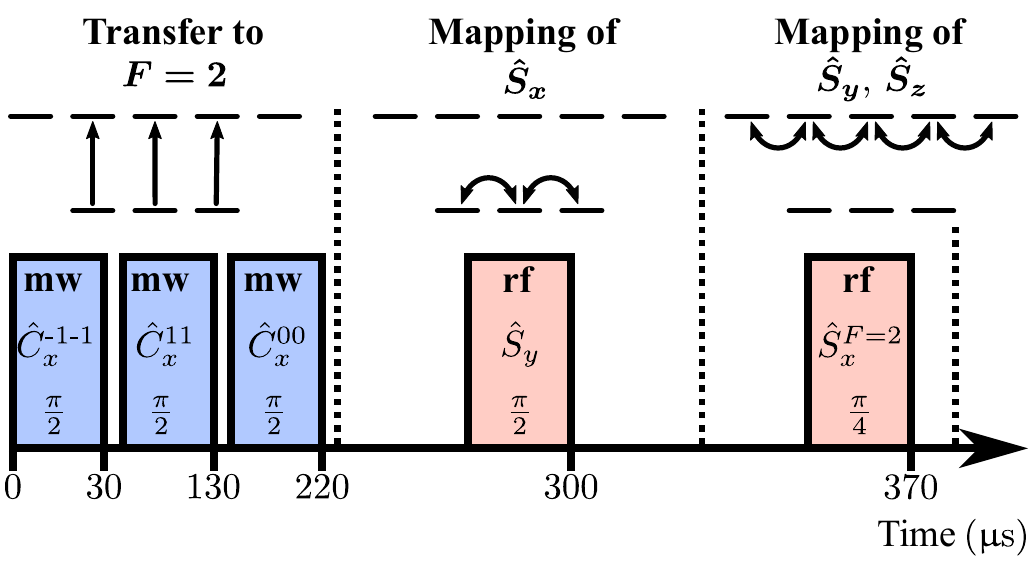}
	\caption{Pulse sequence for the simultaneous readout of all three spin directions. The upper panel depicts the coupling in the two hyperfine manifolds and the boxes contain the rotation operator together with the rotation angle.}
	\label{PulseSeqSpinWave}
\end{figure}
\begin{figure*}
	\centering
	\includegraphics[width=\textwidth]{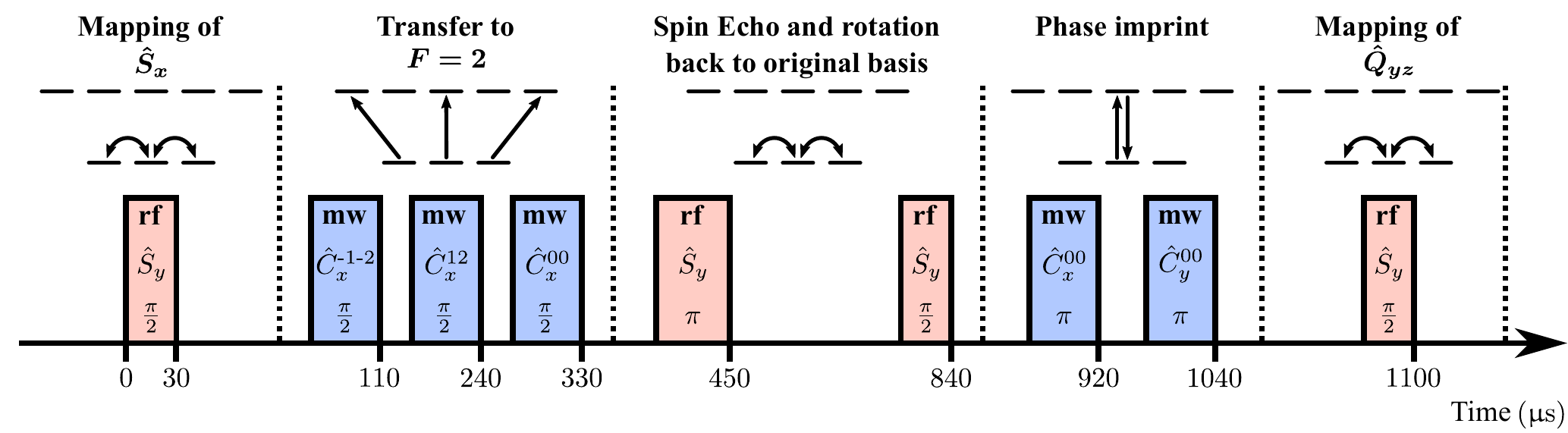}
	\caption{Pulse sequence for the simultaneous readout of $ S_x $ and $ Q_{yz} $. The notation is the same as in Fig.~\ref{PulseSeqSpinWave}.}
	\label{PulseSeqSqueezing}
\end{figure*}
In order to simultaneously read out all three spin directions we apply the pulse sequence as described in the main text which is depicted in Fig.~\ref{PulseSeqSpinWave}.
The total unitary operation describing this measurement protocol reads as follows
\begin{equation}
\hat U = \mathrm{e}^{-i \frac{\pi}{4}\hat{S}_x^{F=2}}\cdot\mathrm{e}^{i \frac{\pi}{2}\hat{S}_y} \cdot\mathrm{e}^{-i\frac{\pi}{2}\hat{C}^{0,0}_x}\cdot\mathrm{e}^{-i\frac{\pi}{2}\hat{C}^{+1,+1}_x}\cdot\mathrm{e}^{-i\frac{\pi}{2}\hat{C}^{-1,-1}_x}.
\end{equation}

Expressing the final projective measurement in the $ F=1 $ and $ F=2 $ manifolds in terms of the original spin-1 states yields the following POVMs:
\begin{equation}
\begin{alignedat}{2}
&\hat E_0 = \hat{a}'^\dagger_{1,-1}\hat{a}'_{1,-1} 	&&=  -\frac{1}{4}\hat{S}_x +\frac{1}{8}\hat{Q}_{xx} + \frac{1}{6}\hat{N}_\text{tot}\\
&\hat E_1 = \hat{a}'^\dagger_{1,0}\hat{a}'_{1,0} 	&&=  -\frac{1}{4}\hat{Q}_{xx} + \frac{1}{6}\hat{N}_\text{tot}\\
&\hat E_2 = \hat{a}'^\dagger_{1,+1}\hat{a}'_{1,+1} 	&&= \frac{1}{4}\hat{S}_x +\frac{1}{8}\hat{Q}_{xx} + \frac{1}{6}\hat{N}_\text{tot} \\
&\hat E_3 = \hat{a}'^\dagger_{2,+2}\hat{a}'_{2,+2} 	&&= \frac{1}{16}\sqrt{\frac{3}{2}}\hat{S}_y + \frac{1}{8\sqrt{2}}\hat{S}_z+\frac{\sqrt{3}}{32}\hat{Q}_{yz}\\
& &&-\frac{1}{32}\hat{Q}_{xx}+\frac{1}{128}\hat{Q}_{zz}+\frac{5}{64}\hat{N}_\text{tot}\\
&\hat E_4 = \hat{a}'^\dagger_{2,+1}\hat{a}'_{2,+1} 	&&= \frac{1}{8}\sqrt{\frac{3}{2}}\hat{S}_y -\frac{1}{16}\hat{Q}_{xx}-\frac{3}{32}\hat{Q}_{zz}\\ & &&+\frac{5}{48}\hat{N}_\text{tot}\\
&\hat E_5 = \hat{a}'^\dagger_{2,0}\hat{a}'_{2,0} 	&&= -\frac{\sqrt{3}}{16}\hat{Q}_{yz}+\frac{3}{16}\hat{Q}_{xx}+\frac{11}{64}\hat{Q}_{zz}\\
& &&+\frac{13}{96}\hat{N}_\text{tot}\\
&\hat E_6 = \hat{a}'^\dagger_{2,-1}\hat{a}'_{2,-1} 	&&= -\frac{1}{8}\sqrt{\frac{3}{2}}\hat{S}_y -\frac{1}{16}\hat{Q}_{xx}-\frac{3}{32}\hat{Q}_{zz}\\ & &&+\frac{5}{48}\hat{N}_\text{tot}\\
&\hat E_7 = \hat{a}'^\dagger_{2,-2}\hat{a}'_{2,-2} 	&&= -\frac{1}{16}\sqrt{\frac{3}{2}}\hat{S}_y - \frac{1}{8\sqrt{2}}\hat{S}_z+\frac{\sqrt{3}}{32}\hat{Q}_{yz}\\
& &&-\frac{1}{32}\hat{Q}_{xx}+\frac{1}{128}\hat{Q}_{zz}+\frac{5}{64}\hat{N}_\text{tot},
\end{alignedat}
\end{equation}
where $ \hat{a}'^\dagger_{F,m_\text{F}}\hat{a}'_{F,m_\text{F}} = U^\dagger \hat{a}^\dagger_{F,m_\text{F}}\hat{a}^{\phantom{\dagger}}_{F,m_\text{F}} U  $ and $ \hat{N}_\text{tot}= \sum_{i=-1}^{1} \hat{a}^\dagger_{1,i}\hat{a}_{1,i}$ the total number which is the normalization of the state.
The measurement maps these observables onto the atom numbers, i.e. ${ \hat{a}'^\dagger_{F,m_\text{F}}\hat{a}'_{F,m_\text{F}}\rightarrow n_{F,m_\text{F}} }$.
From this one can extract the observables of interest by a linear combination of the different $ \hat E_i $.
For example the value of $ S_x $ can be extracted via $ {\hat S_x = 2(\hat E_2-\hat E_0) \rightarrow 2(n_{1,+1}-n_{1,-1})}$.
Note that in the main text we normalize always to the atom number detected in each manifold instead of the total atom number where $ N_1\approx N_2 \approx \frac{1}{2}N_\text{tot} $. Therefore the linear combination of the POVMs to extract the observables of interest differ by a factor of $ 2 $ from the ones in the main text.

\subsection{Simultaneous readout of $ \hat{S}_x $ and $ \hat{Q}_{yz} $}
In order to read out $\hat{S}_x $ and $ \hat{Q}_{yz} $ in a single experimental realization we apply the following pulse sequence as depicted in Fig.~\ref{PulseSeqSqueezing}.
First we apply a resonant rf pulse, corresponding to a $ \pi/2 $-spin rotation around $ \hat{S}_y $, to map the observable $ \hat{S}_x $ onto the population difference of the magnetic substates $ m_\text{F} = \pm1 $ in $ F=1 $.
To "store" the information about the populations during the rest of the readout sequence we use three mw $ \pi/2 $-pulses to transfer half of the population in $(1,0/\pm1) $ to $(2,0/\pm2)  $, respectively.
The states in $ F=2 $ are chosen such that the final Stern-Gerlach maps the states with $ m_\text{F}\neq0 $ onto different spatial positions than the one in $ F=1 $ and thereby allows imaging with a single absorption picture.

These mw pulses require in total $ 350\,\upmu $s.
At this time scale magnetic fluctuations become relevant to which this readout is sensitive.
Therefore, to compensate these fluctuations we use a spin echo technique. For that we apply an rf $ \pi $-pulse and afterwards wait another $ 350\,\upmu $s. 
Note that we calibrated our two rf coils such that the rf-pulses are only resonant with the $ F=1 $ manifold such that we do not change the populations in $ F=2 $.

We then apply another rf $ \pi/2 $-rotation around the $ y $-axis to rotate the state onto its original basis.
To change the readout from $ \hat{S}_x $ to $\hat{Q}_{yz} $ a phase of $ \pi/2 $ has to be imprinted on the state $ m_\text{F} = 0 $. For that we use two resonant $ \pi $-pulses coupling the states $ (1,0)\leftrightarrow(2,0)$. Changing the relative phase of the two pulses by $ \pi/2 $ leads then to the desired phase imprint.
We finally apply another rf $ \pi/2 $-rotation around the $ y $-axis to map the observable $ \hat{Q}_{yz} $ onto the population difference in $ F=1 $.
It is crucial that this last rf pulse is in phase with the first rf pulse as it would otherwise change the measured observable.

\subsection{Preparation of spin waves}
To prepare the spin waves shown in Fig.~\ref{Figure2} as well as for the calibration measurements (next section) we use the following sequences.
After evaporation and loading of the BEC into the elongated dipole trap with trapping frequencies $ (\omega_y,\omega_\perp) = 2\pi\cdot(2.3,170)\, $Hz all atoms are in the state $ (1,-1) $.
Using the rf coils we induce a $ \pi/2 $ spin-rotation around the $ y $-axis. This prepares a homogeneous spin state along the cloud with $ S_x(y)/n_1(y) = 1 $.
By applying a constant current to one of the rf coils we generate a magnetic field gradient with $ \dfrac{\partial B_z}{\partial y}\approx 0.2\,\upmu\text{G}/\upmu\text{m}$ along the longitudinal direction of the cloud.
We employ this gradient for $ 100\, $ms to produce a spin wave with the spin vector rotating  in the $ S_x$$-$$S_y $\,-plane of the spin sphere.
Afterwards we apply another rf spin rotation to tilt the spin wave by $ \approx \pi/4 $.

For calibration we prepare two different spin waves in the elongated dipole trap.
In a first measurement we prepare the spin wave in the $ S_x$$-$$S_y $-plane as described before omitting the last tilting pulse.
In a second measurement we prepare a spin wave in plane parallel to the $ S_x$$-$$S_y $-plane with a finite value of $ S_z/N_\text{tot} = 0.9 $.
This is done by using a shorter rf pulse at the beginning of the spin-wave preparation.

\subsection{Calibration of $ S_x $, $ S_y $ and $ S_z $ readout}
\begin{figure*}
	\centering
	\includegraphics[width=\textwidth]{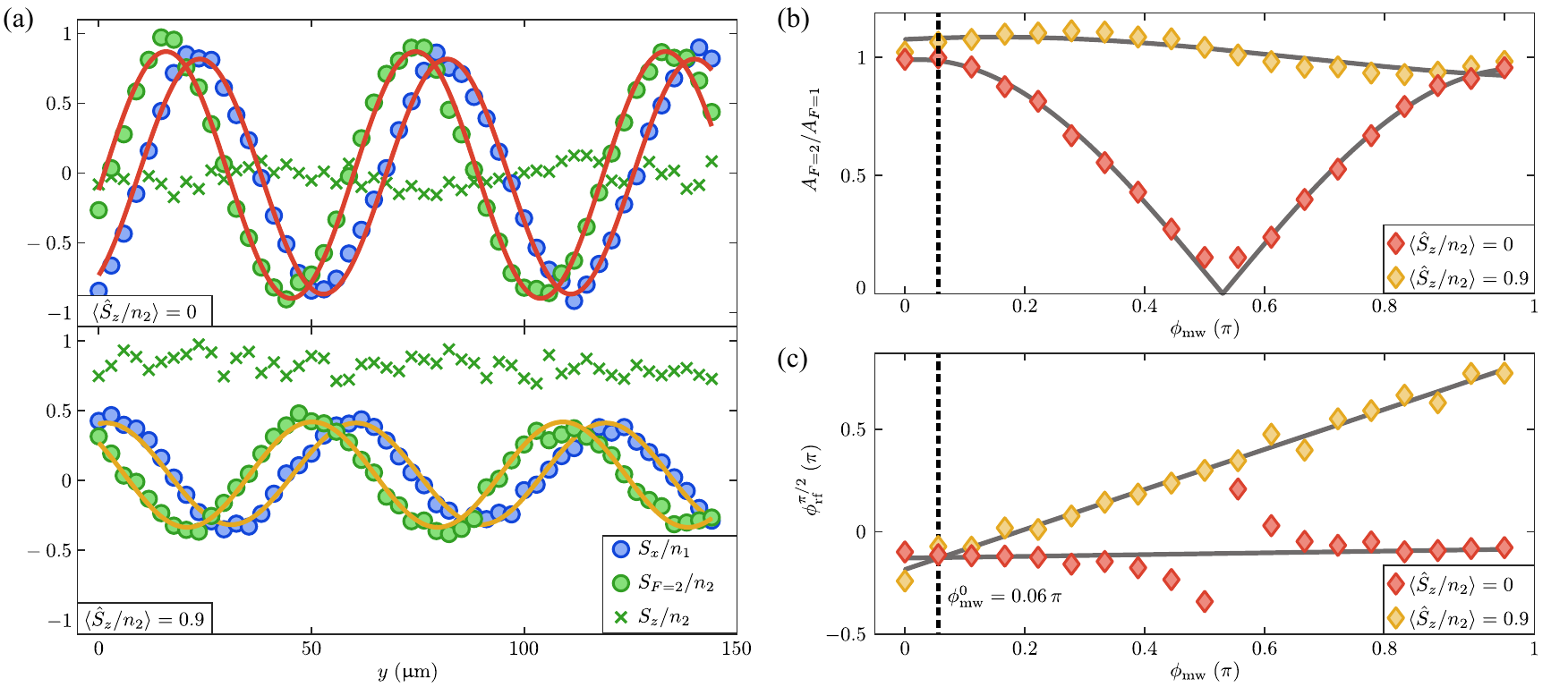}
	\caption{Calibration of the rf and mw phases. (a) For the calibration we prepare two spin waves with constant $ S_z/n_2 = 0  $ (upper panel) and $ S_z/n_2 = 0.9  $ (lower panel) where examples are shown for $ \phi_\text{rf} =  0.1\,\pi$ and $ \phi_\text{rf} = 0 $, respectively.
		We fit the signal extracted from a single absorption image in $ F=1 $ (blue symbols) and $ F=2 $ (green symbols) with a sine to extract their relative spatial phases and amplitudes $ A_F $.
		(b) Amplitudes as a function of $ \phi_\text{mw} $.
		From the spin wave with  $ S_z/n_2 = 0  $ we extract $ \phi_\text{mw}^0  $ from the point where the amplitude ratio is 1.
		(c) Repeating the measurement for fixed $ \phi_\text{mw} $ and different $ \phi_\text{rf} $ we extract the value $ \phi_\text{rf}^{\pi/2} $  at which the two signals have a relative spatial phase shift of $ \pi/2 $.
		We plot the result as a function $ \phi_\text{mw} $.
		Applying a linear fit to both results (gray solid lines) yields the value of $ \phi_\text{mw}^0 = 0.06 \pi$ at the crossing point with a stronger constraint on this parameter than the previous method.\\
	}
	\label{Calibration}
\end{figure*}
For the simultaneous readout of all three spin directions we measure $ S_x $ in $ F=1 $ and $ S_y $ in $ F=2 $ (as depicted in Fig.~\ref{PulseSeqSpinWave}).
The observable that is extracted from $ F=2 $ depends on the relative phases between the final two rf pulses and the mw pulses coupling the two manifolds.
By changing the relative phases of the three mw pulses one adjusts the phases between the states of the $ F=2 $ manifold.
For example a $  $ relative phase  $ \phi_\text{mw}  = \pi/2 $ of the coupling $ (1,0) \leftrightarrow (2,0) $ compared to the other two mw pulses imprints a relative phase of $ \pi/2 $ on the state $ (2,0) $ and therefore changes the final readout in $ F=2 $ from $ \hat{S}_y $ to $ \hat{Q}_{xz} $.
Furthermore, a change in the relative phase between the two rf-pulses $ \phi_\text{rf} $ changes the spin directions that are read out (e.g. $ S_y\rightarrow S_x $).

For calibration of these phases we prepare the two spin waves described above and systematically scan the phases $ \phi_\text{mw} $ and $ \phi_\text{rf} $.
In each measurement, we record the population differences
\begin{align}
\begin{split}
S_x(y) = & \langle\hat{S}_x(y)\rangle_{\delta y} =n_{1,+1}(y) - n_{1,-1}(y) \\
S_y(y) = &\frac{4}{\sqrt{6}}\,(n_{2,+1}(y)-n_{2,-1}(y))\\
S_z(y) = &\sqrt{2}\,[2n_{2,+2}(y)-n_{2,+1}(y) \\
&+ n_{2,-1}(y)-2n_{2,-2}(y)],
\end{split}
\end{align}
where the spin direction detected by $S_{F=2} $ in $ F=2 $ is defined by the reference $ S_x $ in $ F=1 $.

We first employ the spin wave in the $ S_x$$-$$ S_y $-plane.
As expected $ S_{z}/n_2 $ is constant and $ \approx 0 $ (see upper panel Fig.~\ref{Calibration}(a)).
$ S_x/n_1 $ as well as  $ S_{F=2}/n_2 $ are oscillating as a function of position $ y $.
We thus fit a sine to both signals to extract their amplitudes and relative spatial phases.
Changing the phase of the rf-pulse shifts the position of the $ S_{F=2}(y) $ wave with respect to the $ S_x(y) $ wave.
To read out two orthogonal spin directions we extract from this measurement the phase $ \phi^{\pi/2}_\text{rf} $ at which the two waves have a spatial phase shift of $ \pi/2 $ corresponding to $ S_y = S_{F=2}$. 
\begin{figure*}
	\centering
	\includegraphics[width=\textwidth]{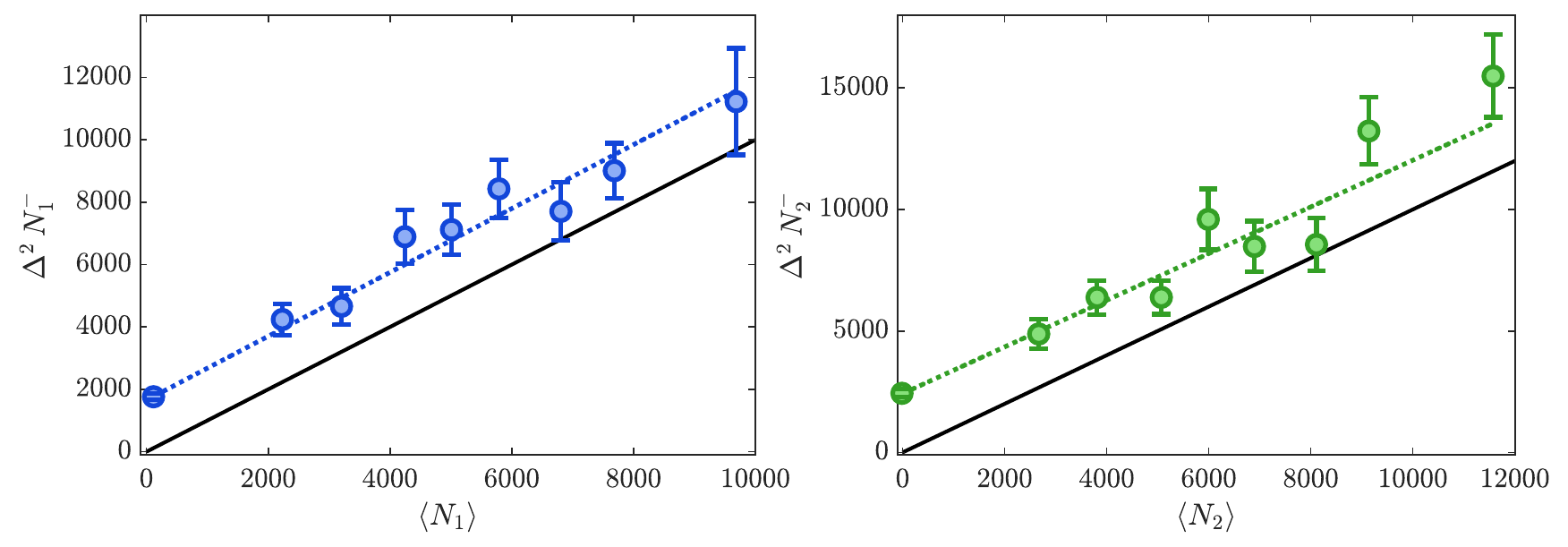}
	\caption{Results of the imaging calibration. The left and right panels show the atom number fluctuations vs the total atom number measured in $ F=1 $ and $ F=2$, respectively. The dashed lines are linear fits to the data while the black lines are the theoretical expectation for a coherent state without technical fluctuations.}
	\label{ImagingCalib}
\end{figure*}

A change in the phase of the mw pulse does not change the value of $ \phi^{\pi/2}_\text{rf} $ but leads to a reduced amplitude of the wave recorded in $ F=2 $ (red points in Fig.~\ref{Calibration}(b)) as this changes the readout from a spin operator in the $ S_x$$-$$S_y $-plane to a quadrupole operator in the $ Q_{xz}$$-$$Q_{yz} $-plane. For the prepared spin wave the mean value of these quadrupole operators is 0. 
Therefore, we can extract the phase value $ \phi^0_\text{mw}  = 0.06\,\pi$ of the mw pulse at which the amplitude of the wave in $ F=2 $ is 1.

As a consistency check, we repeat the calibration with a spin wave with a finite value of $ S_z $.
We use the same readout as before and consistently find a constant value $ S_z/n_2\approx 0.9 $ and an oscillatory behavior for $ S_x $ and  $ S_{F=2} $ (see lower panel Fig.~\ref{Calibration}(a)).
From the latter we extract the phase and amplitude.
As before a change of the phase $ \phi_\text{rf} $ of the rf pulse shifts the relative phase of the two signals from which we extract $ \phi_\text{rf}^{\pi/2} $.
For this wave, however, the mean value of $ Q_{xz}(y) $ and $ Q_{yz}(y) $ are non-vanishing but obey the relation $ Q_{xz} \approx S_{x} $ and $ Q_{yz} \approx S_{y} $.
This means that a change in the phase $ \phi_\text{mw} $ of the mw pulse does not change the amplitude of the signal in $ F=2 $ but its spatial phase.
Thus, the value of $ \phi_\text{rf}^{\pi/2} $ now depends on $ \phi_\text{mw} $ as shown in Fig.~\ref{Calibration}(b).
From the two measurements we then extract the value of $ \phi_\text{mw}^0 $ at which the values of $ \phi_\text{rf}^{\pi/2} $ from both spin waves coincide.
This marks the phase settings to be used in order to measure the two desired spin observables in the two manifolds.

Both methods yield a value of $ \phi_\text{mw}^0\approx 0.06\pi$.
The second method has the advantage that the phase of $ \phi_\text{rf}^{\pi/2}$ depends linearly on $  \phi_\text{mw} $, while the amplitude close to the maximum is quadratic as a function of $ \phi_\text{mw} $.
Therefore the latter allows a more precise calibration of the phase $ \phi_\text{mw}^0 $.
\subsection*{Limitations of the readout scheme}
Here we provide details of how the detected fluctuations are related to the fluctuations of the original state.
In order to see this we consider an ideal $\pi/2$ mw-pulse transferring \emph{on average} half of the atoms to the $F=2$ manifold. 
We want to determine the fluctuations of $S_x$ measured in the $ F=1 $ manifold after the splitting. 
Working in the Heisenberg picture, we have 
\begin{equation}
\hat{S}'_{x} = \frac{1}{\sqrt{2}}\left( \hat{a}_{1,0}'^\dagger(\hat{a}'_{1,1}+ \hat{a}'_{1,-1}) + \text{h.c.}\right).
\end{equation}
To relate back to $\hat S_x$ of the original state before the mw-pulse (which we assume to be described by $\hat C_y^{ii}$), we need to replace $\hat a'_{1,i}=(\hat a_{1,i}+\hat a_{2,i})/\sqrt{2}$, where $a_{2,i}$ are the empty $F=2$ modes. It can be seen that the mean spin in one of the output ports of this beam splitter is half of the spin of the input state. In the case of the prepared squeezed state the mean spin vanishes in all directions. For the variance we thus obtain
\begin{equation}
\begin{split}
&\Delta^2 S'_{x} = \langle (\hat S'_{x})^2 \rangle \\
&= \frac{1}{4}\left\langle\frac{1}{2}\left[ (\hat a_{1,0}^\dagger\!+\!\hat a_{2,0}^\dagger)(\hat a_{1,1}\!+\!\hat a_{2,1}\!+\!\hat a_{1,-1}\!+\!\hat a_{2,-1})\!+\!\text{h.c.}\right]^2 \right\rangle \\
&= \frac{1}{4}\left\langle \hat S_x^2 + \frac{1}{2}\left(2\hat N_{1,0} + (\hat a_{1,1}^\dagger+\hat a_{1,-1}^\dagger)(\hat a_{1,1}+\hat a_{1,-1}) \right)  \right\rangle \\
&\approx  \frac{1}{4}\left(\Delta^2 S_x + N_{\rm{tot}}\right).
\end{split}
\end{equation}
The third line is obtained by exploiting that the $ F=2 $ modes are unoccupied. The "$\approx$" in the last line means that we employ the approximation that the side mode populations are much smaller than $N_{1,0}$.
An analogous calculation can be done for the measurement of $\hat Q_{\rm{yz}}$ after the splitting pulse or for any combination $\hat F(\phi)$ of the two.
We thus recover Eq.~(4) of the main text.

\subsection{Image processing and calibration}
Details about our imaging system and the calibration procedure are reported in \cite{Muessel2013}.
To reduce imaging noise we employ a fringe removal algorithm as detailed in~\cite{Ockeloen2010}.

To check the calibration of our imaging we prepare a coherent spin state with approximately equal mean atom numbers in the states $ (1,\pm1) $ and $ (2,\pm2) $.
Starting from the state $ (1,-1) $ we use two mw pulses coupling the states $ (1,-1)\leftrightarrow(2,0) $ and $ (2,0)\leftrightarrow(1,0) $. For the second pulse we use a fixed $ \pi $-pulse while we vary the length of the first pulse.
Together with a magnetic field gradient (Stern-Gerlach pulse) which expels residual atoms in the magnetic substates $ m_\text{F}\neq0 $ we adjust the total atom number $ N_\text{tot} $ in the state $ (1,0) $.
Subsequently, an rf $ \pi/2 $-pulse is used to prepare an equal superposition of the states $ (1,\pm1) $.
These populations are then again split with two mw $ \pi/2 $-pulses coupling the states $ (1,\pm1)\leftrightarrow(2,\pm2) $.
This preparation leads to an equal probability of $ 1/4 $ to find an atom in one of the four states.

Analogous to the squeezing measurement we divide the atomic signal into two halves and extract the atom number difference $ N_{F}^{-,\text{L/R}} = n^\text{L/R}_{F,+F}-n^\text{L/R}_{F,-F}$ in each half and for each manifold $ F=1,2 $.
To mitigate the technical noise contribution we subtract the value of the right half from the one of the left to obtain $ N_F^{-} = N_{F}^{-,\text{L}} -N_{F}^{-,\text{R}} $.
For each setting of the atom number we compute the variance $ \Delta^2 N_F^{-} $ and plot it vs. the measured mean atom number $ \langle N_F\rangle $ in the respective manifold as shown in Fig.~\ref{ImagingCalib}.
For a coherent state one expects to find multinomial fluctuations of the populations implying $ \Delta^2 N_F^{-} = \langle N_F\rangle $.

From a fit to the data we extract a slope of $ 1.02\pm0.05 $ for $ F=1 $ and a slope of $ 0.96 \pm 0.08 $ for $ F=2 $ which is consistent with coherent state fluctuations.
For the offset we find $ 1,710\pm80 $ for $ F=1 $ and $ 2,090 \pm 170 $ for $ F=2 $. These values include the photon shot noise contribution of $ 1,150 $ for $ F=1 $ and $ 1,490 $ for $ F=2 $ which we compute via Gaussian error propagation from the number of detected photons.

\end{document}